\documentclass[12pt]{iopart}

\begin{document}

\title[Holographic relationships in Lovelock type brane gravity]{Holographic 
relationships in Lovelock type brane gravity}

\author{Efra\'\i n Rojas}

\address{Facultad de F\'\i sica, Universidad Veracruzana, 91000, Xalapa, 
Veracruz, M\'exico}

\ead{efrojas@uv.mx}

\vspace{10pt}


\begin{abstract}
We show that the Lovelock type brane gravity is naturally holographic by 
providing a correspondence between bulk and surface terms that appear in 
the Lovelock-type brane gravity action functional. We prove the existence of 
relationships between the $\mathcal{L}_{\mbox{\tiny bulk}}$ and 
$\mathcal{L}_{\mbox{\tiny surf}}$ allowing $\mathcal{L}_{\mbox{\tiny surf}}$ 
to be determined completely by $\mathcal{L}_{\mbox{\tiny bulk}}$. In the 
same spirit, we provide relationships among the various conserved tensors 
that this theory possesses. We further comment briefly on the 
correspondence between geometric degrees of freedom in both bulk and 
surface space.
\end{abstract}

%
%
%
%
%

\section{Introduction}
\label{sec:1}

Holography is a concept that has received widespread attention 
in a large number of physical theories. Its application covers 
a wide extent of scenarios ranging from optics, condensed matter, 
string theory and gravitation, mainly. In theoretical physics
the most common conception we have is that of a correspondence 
between the dynamics of a system occurring in a $(D-1)$-dimensional 
space (termed \textit{surface}, $\partial \mathcal{M}$) and the 
dynamics in a $D$-dimensional space (termed \textit{bulk}, 
$\mathcal{M}$)~\cite{thooft1993,susskind1995,bousso2002}. In particular, 
for string theory the concept of holography is used to relate two 
different theories: one of them describing the surface while the other 
the bulk. From another point of view, the holography term can be used 
with a slightly different nuance. It is known in gravitation that 
the Einstein-Hilbert action, and its geometric generalization viz. 
the Lanczos-Lovelock gravity action~\cite{lovelock1971}, can be 
splitted into bulk and surface terms closely related each other. This 
fact leads to a correspondence giving account of the \textit{same} 
theory both in the surface and in the bulk which serves to relate the 
degrees of freedom of the same theory either on the bulk or in the 
surface. Following the observation of 
Padmanabhan and 
collaborators~\cite{padmanabhan2006,padmanabhan2006b,padmanabhan2010}, 
this latter equivalence can be considered as a \textit{holograpy of 
the action functional}, avoiding confusion with the notion coming from 
string theory. Certainly, this holographic property of gravitation and its 
deep connection with the thermodynamics of black holes has been 
well exploited~\cite{bekenstein1973,hawking1973,padmanabhan2002}. 

Within the framework of relativistic extended objects, 
often referred to as branes, floating in a background spacetime 
usually treated as fixed, the dynamics is obtained from an action 
constructed by the 
formation of suitable higher-order scalars, using the induced metric 
$g_{ab}$, and the extrinsic curvature tensor $K_{ab}$, which represent the 
geometry of the worldvolumes~\cite{carter1992,carter1995}. 
To have congruence with the mathematical structure, it is required 
that both the worldvolume reparametrization invariance and the diffeomorphisms invariance 
of the background spacetime must be satisfied, which reduces the 
number of possible geometrical scalars. However, a not so pleasant 
feature is that this type of actions yield Euler-Lagrange equations 
involving higher-order derivatives of the field variables. In 
this regard, the Lovelock type brane gravity is a second-order 
theory that involves a set of geometric invariants defined on a 
$(p+1)$-dimensional hypersurface, the trajectory swept out by a 
$p$-dimensional brane, embedded whitin a flat manifold with an 
extra dimension with the particularity that the associated equations 
of motion remain of second-order in the derivatives of the 
fields~\cite{cruzrojas2013,bccrojas2016}. 
This particular aspect makes a given theory free from many of the 
pathologies that plague higher-order derivative theories~\cite{zanelli}. 
This fact is important because it assures no propagation of extra degrees 
of freedom. The underlying common structure which gives rise to this 
theory is that these invariants are polynomials of degree $n \leq p+1$, 
but now in the extrinsic curvature of the hypersurface. The interest 
in this effective field theory is not only for its rich geometric 
structure but also for its possible implications in the description 
of cosmological acceleration behaviors in the brane-like universe 
scenarios~\cite{dgp2000,rham2010,trodden2011,trodden2011b}. These Lovelock brane 
invariants are similar in form either to the original Lanczos-Lovelock 
invariants in pure gravity or to their necessary counterterms in 
order to have a well posed variational problem~\cite{lovelock1971,myers87,davis03,olea05}. 
This analogy needs a word of caution. While for even values of $n$ 
the Lovelock brane invariants look like the Gauss-Bonnet invariants, 
for odd values of $n$ the corresponding Lovelock brane invariants 
acquire the form of the Gibbons-Hawking-York-Myers boundary 
terms~\cite{myers87,padmanabhan2017} which are seen as counterterms 
if we have the presence of bulk Lovelock invariants. Unlike what 
happens with the counter-terms either in the Einstein theory or 
more generally in the Lanczos-Lovelock gravity, for 
the odd terms in this type of gravity we have the presence of time 
derivatives of the field variables, which is a sign that we have 
dynamics on this type of hypersurfaces. This important fact leads to a handful 
of attractive features. 

In this spirit, the holographic scheme of our interest considers that 
from a second-order Lagrangian density $\mathcal{L}$ we can establish a 
splitting of it into two parts; the first being a first-order derivative
Lagrangian density, $\mathcal{L}_{\mbox{\tiny bulk}}$, and the second 
a divergence absorbing the second-order derivative content, 
$\mathcal{L}_{\mbox{\tiny surf}}$, with the attribute of being able to establish 
a specific relationship between these such that it allows to determine 
$\mathcal{L}_{\mbox{\tiny surf}}$ completely in terms of 
$\mathcal{L}_{\mbox{\tiny bulk}}$. This type of correspondences were 
termed also as \textit{holographic relationships}~\cite{padmanabhan2006}. 
With this understanding, the bulk and the surface terms in Lovelock 
type brane gravity must encode the same amount of dynamical content. 
Within the quantum framework of brane-like 
universes~\cite{davidson1998,davidson1999,davidson2003,ostro2009,modified2012,qmodified2014} 
and the Dirac's extensible model for the 
electron~\cite{onder1988a,onder1988,relectron2011,biswajit2013,turcos2019} the usage of this 
holographic concept has been used in order to extract information from 
one term of the splitted Lagrangian, based in the other.

This paper is devoted to show that the Lovelock type brane gravity 
also possesses a holographic nature by providing the existence of 
holographic relationships. To some extent this fact is reasonable 
since this type of gravity belongs to the set of theories known as 
affine in acceleration, i.e., linear in second order derivatives of 
the fields~\cite{affine2016}, allowing the identification of a 
divergence term. In our approach, we have a larger number of 
holographic relationships since we have at most the double of the 
geometric invariants in comparison with the original Lanczos-Lovelock gravity.  
Additionally, we highlight the role played by the different conserved 
tensors that this theory possesses in our development as well as the 
relationships among them. Moreover, it will be interesting to explore 
whether these holographic relationships may also connect black hole 
solutions with their termodinamical properties in the braneworld 
scenario once suitable geometries be adapted, as occur in the case 
of the original Lovelock gravity theory. For some of the black holes 
solutions in this direction see~\cite{hawking2000,dadhich2000,majumdar2005}.

The remainder of the paper is organized as follows. In section 2 
we provide an overview of the Lovelock theory for extended objects 
geodesically floating in a flat Minkowski spacetime. We derive holographic 
relationships for the Lovelock type brane gravity in section 3. In 
section 4 we provide some relationships among the different conserved 
tensors of this theory. In section 5 we provide 
a discussion about the splitting into two parts of the first Lovelock
brane invariant, named $K$ brane action. Conclusions and comments
are presented in section 6. Throughout our analysis 
the convention for the worldvolume Riemann tensor we follow is 
$\mathcal{R}_{abc}{}^d = - 2 \partial_{[a} \Gamma^d _{b]c} 
+ 2\Gamma^e_{c[a}\Gamma^d_{b]e}$ where $T^{[ab]}$ indicates 
anti-symmetrization under the convention $T^{[ab]} = (T^{ab} 
- T^{ba})/2$. Similarly, $T^{(ab)}$ indicates symmetrization according to 
$T^{[ab]} = (T^{ab} + T^{ba})/2$. Somewhat larger computations were put in an Appendix.

\section{Lovelock type brane theory}
\label{sec:2}

The system of interest is a $p$-dimensional spacelike extended object, $\Sigma$, 
geodesically floating in a $N=(p+2)$-dimensional fixed Minkowski spacetime ${\cal M}$ 
with metric $\eta_{\mu \nu}$ $(\mu,\nu =0,1,\ldots, p+1)$. To specify the $\Sigma$ trajectory,
known as worldvolume and denoted by $m$, we set $y^\mu = X^\mu(x^a)$ where 
$y^\mu$ are local coordinates of ${\cal M}$ and  $x^a$ are local coordinates 
of $m$, being $X^\mu$ the embedding functions $(a,b= 0,1,\ldots,p)$.
Within the geometric framework of extended objects the essential derivatives 
of $X^\mu$ enter the game through the induced metric tensor $g_{ab} = 
\eta_{\mu \nu} e^\mu{}_a e^\nu{}_b$ and the extrinsic curvature $K_{ab} = - \eta_{\mu\nu}n^\mu 
 \nabla_a e^\nu{}_b = K_{ba}$ where $e^\mu{}_a = \partial_a X^\mu$ are the 
tangent vectors to $m$, $n^\mu$ is the spacelike unit normal vector to $m$, 
and $\nabla_a$ is the worldvolume covariant derivative, $\nabla_a g_{bc}=0$.

We are interested in the Lovelock type brane gravity theory~\cite{cruzrojas2013}. 
For a $(p+1)$-dimensional worldvolume described parametrically by the field
variables $X^\mu$, the action functional
\begin{equation}
S[X^\mu] =  \int_m d^{p+1} x \, \sqrt{-g} \sum_{n=0} ^{p+1} \alpha_n\,L_n (g_{ab},
K_{ab}),
\label{eq:Lbaction}
\end{equation}
where
\begin{equation}
L_n (g_{ab}, K_{ab}) = \delta^{a_1 a_2 a_3 \cdots a_n} _{b_1 b_2 b_3 \cdots b_n}
 K^{b_1}{}_{a_1} K^{b_2}{}_{a_2} K^{b_3}{}_{a_3} \cdots K^{b_n}{}_{a_n},
\label{eq:lovelock-brane}
\end{equation}
ensures that the Euler-Lagrange equations for the field variables are of second
order. Here,  $\alpha_n$ are constants with appropriate dimensions, $g 
:= \det (g_{ab})$ and $\delta^{a_1 a_2 a_3 \ldots a_n} _{b_1 b_2 \, b_3 \ldots \, b_n}$ 
denotes the alternating tensor known as the generalized Kronecker delta (gKd). 
Certainly, taking as a guideline the original Lovelock theory of gravity, the 
action~(\ref{eq:Lbaction}) has been formed but now considering the antisymmetric 
products of the extrinsic curvature $K^a{}_b$. Written out in full, the gKd is 
given by the determinant made of Kronecker delta functions
\begin{equation}
 \delta^{a_1 a_2  \ldots a_{n-1} a_n} _{b_1 b_2 \ldots b_{n-1} b_n}
= 
\left|
\matrix{
\delta^{a_1} _{b_1} & \delta^{a_1} _{b_2} & \ldots & 
\delta^{a_1} _{b_{n-1}} & \delta^{a_1}_{b_n} 
\cr
\delta^{a_2} _{b_1} & \delta^{a_2} _{b_2} & \ldots & 
\delta^{a_2} _{b_{n-1}} & \delta^{a_2}_{b_n} 
\cr
\vdots & \vdots & \ddots & \vdots & \vdots
\cr
\delta^{a_{n-1}} _{b_1} & \delta^{a_{n-1}} _{b_2} & \ldots & 
\delta^{a_{n-1}} _{b_{n-1}} & \delta^{a_{n-1}}_{b_n} 
\cr
\delta^{a_n} _{b_1} & \delta^{a_n} _{b_2} & \ldots & 
\delta^{a_n} _{b_{n-1}} & \delta^{a_{n}} _{b_n}
}
\right|.
\end{equation}
Regarding the definition~(\ref{eq:lovelock-brane}), we set $L_0 = 1$. 
We must to stress that the action functional~(\ref{eq:Lbaction}) is 
invariant under reparametrizations of the worldvolume. In view of 
the fact that the Lagrangian~(\ref{eq:lovelock-brane}) is a 
polynomial of degree $n \leq p+1$ in the extrinsic curvature, 
the action~(\ref{eq:Lbaction}) is a second-order derivative theory. The 
geometrical invariants~(\ref{eq:lovelock-brane}) are known as 
\textit{Lovelock brane invariants} (LBI). By construction, these 
terms vanish for $n>p+1$ whereas the term with $n= p+1$ corresponds 
to a topological invariant not contributing to the field equations. 
Since the independent variables to describe the worldvolume are the 
embedding functions instead of the metric, we then have one greater 
number of Lovelock type brane terms contrary to the pure gravity case. 
For even $n$ we recognize the form of the Gauss-Bonnet (GB) terms but 
expressed now in terms of the worldvolume geometry; for $n=0$ we 
have the DNG Lagrangian, for $n=2$ we have the Regge-Teitelboim (RT) 
model~\cite{rt,tapia1989,pavsic2001,paston2010,davidson1998,davidson2003,ostro2009,modified2012}, 
for $n=4$ we have the standard GB Lagrangian which for $p>3$ produces 
non-vanishing equations of motion with ghost-free 
contribution \cite{trodden2011b,germani2002,maeda2004}. On 
the other side, for odd $n$ the corresponding LBI are seen as the 
Gibbons-Hawking-York-Myers boundary terms which may exist if we have 
the presence of bulk Lovelock 
invariants~(see~\cite{lovelock1971,cruzrojas2013} for further details). 
In short, for a $p$-brane there are at most $p+1$ possible terms 
leading to second-order equations of motion~\cite{cruzrojas2013,bccrojas2016}.

Given that the background spacetime is flat Minkowski, in order to 
have a surface that propagates here, both intrinsic and extrinsic 
geometries of the surface can not be chosen arbitrarily but must 
satisfy the Gauss-Codazzi and Codazzi-Mainardi integrability conditions 
given by ${\cal R}_{abcd} = K_{ac} K_{bd} - K_{ad}K_{bc}$ and $\nabla_a 
K_{bc} - \nabla_b K_{ac} = 0$, respectively. Here, ${\cal R}_{abcd}$ denotes 
the worldvolume Riemann tensor. In this sense, the repeated application 
of the Gauss-Codazzi condition in~(\ref{eq:lovelock-brane}), gives rise 
to express the LBI for the even and odd cases in terms of 
${\cal R}^a{}_{bcd}$. Indeed, for $n=0,1,2,3,\ldots$ we have
\begin{eqnarray}
L_{(2n)} &= \frac{1}{2^n} \delta^{a_1 a_2 \cdots a_{2n-1} 
a_{2n}}_{b_1 b_2 \cdots b_{2n-1} b_{2n}} \mathcal{R}^{b_1b_2}{}_{a_1 a_2} 
\cdots \mathcal{R}^{b_{2n -1} b_{2n}}{}_{a_{2n-1} a_{2n}}, 
\label{evenL}
\\
L_{(2n + 1)} &= \frac{1}{2^n} \delta^{a_1 a_2 \cdots a_{2n} 
a_{2n +1}}_{b_1 b_2 \cdots b_{2n} b_{2n +1}} \mathcal{R}^{b_1b_2}{}_{a_1 
a_2} \cdots \mathcal{R}^{b_{2n -1}
b_{2n}}{}_{a_{2n-1} a_{2n}} K^{b_{2n+1}}{}_{a_{2n+1}}.
\label{oddL}
\end{eqnarray}
These relations will prove useful to carry out the computations 
leading to holographic relationships in this type of gravity. Indeed, 
when the LBI are expressed in this fashion, it may be suggested 
to consider alternative sets of independent variables instead of $X^\mu$.

In view of the importance of some conserved tensors inherent in 
this theory, as well as to get a better understanding of our 
development, we glance at the manifestly covariant 
variation approach of the action in order to obtain the extrema 
conditions of the model (for more details see~\cite{cruzrojas2013,bccrojas2016}). 
We perform the variation of the action~(\ref{eq:Lbaction}) in steps. We start with
\begin{equation}
\delta S = \int_m d^{p+1} x \,\left[ \delta (\sqrt{-g}) 
\sum_{n=0}^{p+1} \alpha_n L_n + \sqrt{-g} \sum_{n=0}^{p+1} \alpha_n 
\delta L_n \right].
\label{var1}
\end{equation}
Now, we must recall the well known expression for 
the first variation, $\delta (\sqrt{-g}) = (\sqrt{-g} /2) g^{ab} 
\delta g_{ab}$. Afterwards, from~(\ref{eq:lovelock-brane}), 
the second variation yields
\begin{eqnarray}
\delta L_n &=& n \,\delta^{a_1 a_2 a_3 \cdots a_n}_{b_1 b_2 b_3 
\cdots b_n} K^{b_2}{}_{a_2} K^{b_3}{}_{a_3} \cdots K^{b_n}{}_{a_n}
\delta K^{b_1}{}_{a_1},
\nonumber
\\
&=& n \, J^a_{(n-1)\,b} \delta K^b{}_a,
\label{var2}
\end{eqnarray}
where have introduced the worldvolume tensor~\cite{cruzrojas2013}
\begin{equation}
\label{eq:conserved1}
J^{a}_{(n)b} := \delta^{a a_1 a_2 a_3 \ldots a_n} _{b b_1 b_2 b_3 
\ldots b_n} K^{b_1}{}_{a_1} K^{b_2}{}_{a_2} K^{b_3}{}_{a_3} \ldots 
K^{b_n}{}_{a_n},
\end{equation}
for the $n$th order Lovelock type brane invariant. Additionally, 
$\delta K^a{}_b = \delta (g^{ac} K_{cb}) = - g^{ad} K^c{}_b \delta 
g_{dc} + g^{ac} \delta K_{cb}$. Thence, the variation~(\ref{var2})
becomes $\delta L_n = \left( J^{ab}_{(n)} - g^{ab} L_n \right)
\delta g_{ab} + n \, J^{ab} _{(n-1)} \delta K_{ab}$, in which, we
have considered that the tensor~(\ref{eq:conserved1}) obeys the useful
recurrence relation $J^a_{(n)b} = \delta^a_b\,L_n - n K^a{}_c J^c_{(n-1)b}$,
\cite{cruzrojas2013,bccrojas2016}.
Thus, the variation~(\ref{var1}) can be written in the form
\begin{equation}
\label{var3}
\delta S = \int_m d^{p+1}x\, \sqrt{-g} \sum_{n=0}^{p+1} \alpha_n
\left[ \left( J^{ab}_{(n)} - \frac{1}{2} g^{ab} L_n \right) \delta g_{ab}
+ n\,J^{ab} _{(n-1)} \delta K_{ab} \right].
\end{equation}
Some important remarks, before continuing, are in order. The 
tensors~(\ref{eq:conserved1}) are symmetric since they inherit 
the symmetries from the extrinsic curvature. These are conserved, 
$\nabla_a J^{ab} _{(n)} = 0$, which may readily be proved by 
using the Codazzi-Mainardi integrability condition for extended 
objects, $\nabla_a K_{bc} = \nabla_b K_{ac}$.
The term \textit{Lovelock brane tensors}, for 
expression~(\ref{eq:conserved1}), has been coined to a 
large extent because  they play a similar role to those conserved 
tensors appearing in the pure Lanczos-Lovelock gravity. For a 
$(p+1)$-dimensional worldvolume there are at most an equal number 
of conserved tensors $J^{ab} _{(n)}$. Finally, it is fairly direct to
verify that $J^{ab}_{(n)} = (1/n) (\partial L_n/ \partial K_{ab})$, for $n\neq 0$. 

Let us return to the variation~(\ref{var3}). Since the field variables
are the functions $X^\mu$, the response of the action (\ref{eq:Lbaction})
to small changes in the worldvolume, $X^\mu (x^a) \rightarrow X^\mu (x^a)
+ \delta X^\mu (x^a)$, is done through the variations $\delta g_{ab}$
and $\delta K_{ab}$. The deformation $\delta X^\mu$ can be decomposed 
into its parts, normal and tangential to the worldvolume, $\delta X^\mu
= \Phi\,n^\mu + \Phi^a \,e^\mu{}_a$. The  motions transverse to the 
worldvolume are the physically relevant ones so the common strategy is 
consider only that $\delta_\perp X^\mu = \Phi n^\mu$ where $\Phi = \eta_{\mu\nu}
n^\mu \delta X^\nu$ it is 
assumed small in addition to be an arbitrary function of $x^a$. 
The tangential deformation can be identified with a reparametrization of 
the worldvolume, so that $\delta_\parallel S$ contributes only a total 
derivative, $\delta_\parallel S = \nabla_a (\Phi^a\,S)$. Taking into account
both types of deformations, $g_{ab}$ and $K_{ab}$ respond as~\cite{capo1995}
\begin{eqnarray}
\delta g_{ab} &=& 2 K_{ab} \,\Phi + 2 \nabla_{(a} \Phi_{b)},
\label{var4}
\\
\delta K_{ab} &=& - \nabla_a \nabla_b \Phi + K_{ac} K^c{}_b \,\Phi
+ \Phi^c \nabla_c K_{ab} + 2 K_{c(a} \nabla_{b)} \Phi^c.
\label{var5}
\end{eqnarray}
After substituting these into~(\ref{var3}), a lengthy but straightforward computation leads to $\delta S$ to take the form
\begin{equation}
\delta S = \int_m d^{p+1}x\,\sqrt{-g} \left[ \mathcal{E} (L) n_\mu \,
\delta X^\mu + \int_m d^{p+1}x\,\sqrt{-g} \,
\nabla_a Q^a \right],
\label{var6}
\end{equation}
where
\begin{equation}
\mathcal{E} = \sum_{n=0}^{p+1} \alpha_n J^{ab}_{(n)} K_{ab} = 0, 
\end{equation}
is the Euler-Lagrange derivative which in turn can be expressed in terms
of the LBI. Further, $Q^a$ denotes the Noether current  given by
the worldvolume vector
\begin{equation}
Q^a := \sum_{n= 0}^{p+1} \alpha_n \left( J^{ab}_{(n)} e_{\mu\,b}
\,\delta X^\mu - n \,J^{ab}_{(n-1)} n_\mu  \,\delta \nabla_b X^\mu \right).
\label{Qa}
\end{equation}
It should be pointed out that we have only one equation of 
motion which is second-order in the field variables. This 
is related to the fact that we have only one relevant degree 
of freedom for this type of branes which is associated with 
the geometric configuration itself of the worldvolume. Certainly, 
as already mentioned, the physically observable measure of the 
deformation of the worldvolume, is the breathing mode provided 
by the scalar field $\Phi$,~\cite{bccrojas2016}.

Following a similar strategy to the one made to obtain~(\ref{evenL}) 
and~(\ref{oddL}), we have a close relationship between the $J^{ab}_{(n)}$ 
and the worldvolume Riemann tensor. A repeated application of the 
Gauss-Codazzi integrability condition in the 
definition~(\ref{eq:conserved1}) yields the handy identities 
\begin{eqnarray}
\fl J^a_{(2n)b} &= \frac{1}{2^n} \delta^{a a_1 a_2 \cdots a_{2n-1} 
a_{2n}}_{b b_1 b_2 \cdots b_{2n-1} b_{2n}} \mathcal{R}^{b_1 
b_2}{}_{a_1 a_2} \cdots \mathcal{R}^{b_{2n-1}b_{2n}}{}_{a_{2n-1} a_{2n}},
\label{J2n}
\\
\fl J^a_{(2n+1)b} &= \frac{1}{2^n} \delta^{a a_1 a_2 \cdots a_{2n-1} a_{2n}
a_{2n+1}}_{b b_1 b_2 \cdots b_{2n-1} b_{2n} b_{2n+1}} 
\mathcal{R}^{b_1 b_2}{}_{a_1 a_2} \cdots \mathcal{R}^{b_{2n-1} b_{2n}}{}_{a_{2n-1}a_{2n}}
K^{b_{2n+1}}{}_{a_{2n+1}},
\label{J2n+1}
\end{eqnarray}
for  $n=0,1,2, 3, \ldots$.

The framework outlined above serves to argue that, except for 
the $K$ term, all the LBI satisfy the structure
\begin{equation}
 \sqrt{-g} L= \sqrt{-g} \,Q_{a}{}^{bcd} \mathcal{R}^a{}_{bcd},
\end{equation}
where $Q_{a}{}^{bcd}$ is a tensor that has all the symmetries of 
the worldvolume Riemann tensor, made from the induced metric, the 
Riemann tensor and the extrinsic curvature itself, besides a zero 
divergence, $\nabla_c Q_{a}{}^{bcd} = 0$. This will be proved shortly.

In passing we remark that, due to the fact that the worldvolume 
is an oriented timelike manifold embedded in $\mathcal{M}$, the 
odd Lovelock type brane invariants, unlike what happens with the 
common counter-terms in the pure Lovelock gravity, they have the 
presence of time derivatives, which indicates that we have dynamical 
content on this type of hypersurfaces \cite{modified2012,qmodified2014,onder1988a,onder1988,relectron2011,biswajit2013,turcos2019}. 
It should also be noted that there are formal descriptions in 
which some hypersurfaces are considered to play the role of boundaries for gravitational spaces, and which are described by actions that represent 
counter-terms of the original action. Such hypersurfaces can 
have a timelike, spacelike or null-like causal 
structure~\cite{padmanabhan2016,myers2016}. 

\section{Holographic relationships}

The way in which the geometric invariants~(\ref{eq:lovelock-brane})  
were splitted into (\ref{evenL}) and~(\ref{oddL}) becomes important 
because, except for the $K$ brane term, any $n$th Lovelock type 
brane density acquires the form
\begin{equation}
\label{dens1}
\mathcal{L}_n = \sqrt{-g}\, _{(n)}\mathcal{Q}_a{}^{bcd} 
\mathcal{R}^a{}_{bcd},
\end{equation}
where $_{(n)}\mathcal{Q}_a{}^{bcd}$ is a tensor with specific properties. 
We turn now to prove that any $n$th order Lovelock type brane model allows 
a decomposition in terms of both a bulk and surface components of the action 
functional in which both terms are directly related. To do this we will proceed 
in two parts.

\subsection{Even case}

Clearly,~(\ref{evenL}) furnishes us with the structure given 
in~(\ref{dens1}) for $n=1,2,3,\ldots,$
\begin{equation}
L_{(2n)}  = \,  _{(2n)}Q_{b_{2n-1}}{}^{a \,a_{2n-1}a_{2n}}\,
\mathcal{R}^{b_{2n -1}}{}_{a\,a_{2n-1} a_{2n}},
\label{eq:evenL}
\end{equation}
with
\begin{equation}
\fl \, _{(2n)}Q_{b_{2n-1}}{}^{a \,a_{2n-1} a_{2n}} := \frac{1}{2^n} 
\delta^{a_1 a_2 \cdots a_{2n-1} a_{2n}}_{b_1 b_2 \cdots b_{2n-1} 
b_{2n}} \,g^{a \,b_{2n}}\,\mathcal{R}^{b_1b_2}{}_{a_1 a_2} \cdots 
\mathcal{R}^{b_{2n-3} b_{2n-2}}{}_{a_{2n-3} a_{2n - 2}}.
\label{Q1}
\end{equation}
With the indices conveniently placed, it is evident that this tensor 
inherits the symmetries from the Riemann tensor
\begin{equation}
  _{(2n)}Q^{abcd} = -  _{(2n)}Q^{abdc} = - _{(2n)}Q^{bacd} 
= \, _{(2n)}Q^{cdab},
\label{symm1}
\end{equation}
and it is constructed from $g^{ab}$ and $\mathcal{R}^a{}_{bcd}$ or, 
from a brane point of view, from $g^{ab}$ and $K_{ab}$ when the 
worldvolume Riemann tensor is expressed in terms of the extrinsic 
curvature via the Gauss-Codazzi equations for the worldvolume. 
The first values of this $Q$ tensor are
\begin{eqnarray}
\, _{(2)}Q^{abcd} &= g^{a[c} g^{d]b} = J^{a[c}_{(0)} g^{d]b},
\label{Q20}
\\
\, _{(4)}Q^{abcd} &= \mathcal{R}^{abcd} - 2 G^{a[c}g^{d]b}
- 2g^{a[c} \mathcal{R}^{d]b},
\nonumber
\\
&= J^{a[c}_{(2)} g^{d]b} - 2 g^{a[c} J^{d]e}_{(1)} K^b{}_e
+ \mathcal{R}^{abcd},
\label{Q40}
\end{eqnarray}
with $G^{ab} = \mathcal{R}^{ab} - (\mathcal{R}/2) g^{ab}$
being the worldvolume Einstein tensor, and $J^{ab}_{(0)}, J^{ab}_{(1)}$
and $J^{ab}_{(2)}$ are the first Lovelock brane tensors
given by~(\ref{eq:conserved1}),~\cite{cruzrojas2013}. 

Consequently, relationship~(\ref{eq:evenL}) allows us to express 
the Lagrangian densities for the even case as
\begin{equation}
 \mathcal{L}_{(2n)} = \sqrt{-g}\,  _{(2n)}Q_{a}{}^{bcd}\,
\mathcal{R}^{a}{}_{bcd}.
\label{dens2}
\end{equation}
With the aid of the Riemann tensor, $\mathcal{R}_{abc}{}^d = - 2 
\partial_{[a} \Gamma^d _{b]c} + 2\Gamma^e_{c[a}\Gamma^d_{b]e}$, 
the former equation can be written in the form 
\begin{eqnarray}
\nonumber
 \mathcal{L}_{(2n)} &=& \partial_c \left( 2\sqrt{-g}
 \, _{(2n)}Q_{a}{}^{bcd}\Gamma^{a}_{d b} \right) 
 - 2 \left( \partial_{c} \sqrt{-g} \right)
\, _{(2n)}Q_{a}{}^{bcd}\,\Gamma^{a}_{d b}
\\
 &-& 2\sqrt{-g}\left( \partial_{c} \, _{(2n)}Q_{a}{}^{bcd} \right)
\Gamma^{a}_{d b} - 2 \sqrt{-g}\,  _{(2n)}Q_{a}{}^{bcd} 
\Gamma^e _{bc} \Gamma^{a}_{de}.
\nonumber
\end{eqnarray}
Now, by virtue of the identity $\partial_{a} \sqrt{-g} = 
\sqrt{-g} \,\Gamma^{b}_{b a}$, we have
\begin{eqnarray}
\mathcal{L}_{(2n)} &= \partial_{c}\left( 2\sqrt{-g}\,
 _{(2n)}Q_{a}{}^{bcd} \Gamma^{a}_{bd} \right) + 2 \sqrt{-g}
 \,  _{(2n)}Q_{a}{}^{bcd} \Gamma^{a} _{de} \Gamma^e _{bc} 
\nonumber
\\
&- 2 \sqrt{-g}\left( \nabla_{c} \, _{(2n)}Q_{a}{}^{bcd} \right)
\Gamma^{a}_{b d}.  
\nonumber
\end{eqnarray}
It follows therefore that, as long as the condition 
$\nabla_{c} \left[ \, _{(2n)}Q_{a}{}^{bcd} \right] =0$ 
is satisfied, the previous equation takes the form
\begin{equation}
\mathcal{L}_{(2n)} = \mathcal{L}_{(2n)\,\mbox{\tiny bulk}} + 
\mathcal{L}_{(2n)\,\mbox{\tiny sur}},
\label{holo1}
\end{equation}
where
\begin{eqnarray}
\mathcal{L}_{(2n)\,\mbox{\tiny bulk}} & = 2 \sqrt{-g}
\,  _{(2n)}Q_{a}{}^{bcd} \Gamma^{a} _{de} \Gamma^e _{bc},
\label{L2}
\\
\mathcal{L}_{(2n)\,\mbox{\tiny sur}} & = \partial_{c}
\left( 2\sqrt{-g}\, _{(2n)}Q_{a}{}^{bcd} \Gamma^{a}_{b d} \right).
\label{L1}
\end{eqnarray}
Some aspects of this framework are in order. It should be notice 
that for the RT model~(\ref{L2}) is quadratic in $\Gamma^a_{bc}$; 
this evident  by using~(\ref{Q20}). In this approach the form of 
$\mathcal{R}_{abc}{}^d$, expressed entirely in terms of the 
Christoffel symbols and its derivatives without requiring 
$g^{ab}$ and $K_{ab}$, has been an important ingredient in 
arriving to~(\ref{holo1}). 
This fact, allows the decomposition of the Lagrangian densities 
in terms of a surface  and bulk terms. Relation~(\ref{L2}) can 
be considered as a variant of the so-called Dirac-Schr\"{o}dinger 
Lagrangian, or $\Gamma \Gamma$ Lagrangian for short, for Einstein 
theory~\cite{padmanabhan2002b}. According to the symmetries~(\ref{symm1}) 
the divergence of the tensor $Q$ on any of the indices vanishes. This 
is proved in~\ref{app1a}.

It is necessary to stress that $Q_a{}^{bcd} = Q_a{}^{bcd} (g^{ab}, 
\mathcal{R}^a{}_{bcd})$ so the differentiation of~(\ref{L2}) with 
respect to the Christoffel symbols infers that
\begin{equation}
\nonumber
\delta^r _t \frac{\partial \mathcal{L}_{2n\,\mbox{\tiny 
bulk}}}{\partial \Gamma^r_{st}} = 2 \sqrt{-g} \left[ \,  
_{(2n)}Q_{r}{}^{bcs} \Gamma^{r} _{bc} + \,  _{(2n)}Q_{a}{}^{srd} 
\Gamma^{a} _{dr} \right],
\label{partial}
\end{equation}
where $\partial \Gamma^a_{bc} / \partial \Gamma^r_{st} = 
\delta^a _r \delta^s _b \delta^t _c$ has been considered.
Due to the symmetry properties of the tensor~(\ref{Q1}), it is 
necessary that the second term of this expression should vanish. 
Clearly, comparison of the former expression with 
relation~(\ref{L1}) furnishes us with a  holographic relation
\begin{equation}
\label{holographic1}
\mathcal{L}_{(2n)\,\mbox{\tiny sur}} = - \partial_a
\left( \delta^c_b \, \frac{\partial 
\mathcal{L}_{(2n)\,\mbox{\tiny bulk}}}{\partial \Gamma^c_{ab}}  \right).
\end{equation}
We find then that~(\ref{holo1}), with the aid of the 
identity~(\ref{holographic1}), 
can be expressed as
\begin{equation}
\mathcal{L}_{(2n)} = \mathcal{L}_{(2n)\,\mbox{\tiny bulk}}
- \partial_a \left( \delta^c_b \, \frac{\partial 
\mathcal{L}_{(2n)\,\mbox{\tiny bulk}}}{\partial \Gamma^c_{ab}} \right).
\label{holographic11}
\end{equation}
This way of approaching the problem is similar in the spirit to the 
one carried out in~\cite{padmanabhan2006} for the case of pure 
Lovelock gravity. It remains to verify that $\nabla_{c} \left[
\,_{(2n)}Q_{a}{}^{bcd} \right] =0$.  In fact, on geometrical grounds, the 
the invariance under reparametrizations of the worldvolumes in this
type of gravity framework requires this condition. 

Alternatively, there is another form of express the holographic 
relation~(\ref{holographic1}). By using the well known relation 
for the Christoffel symbols in terms of the geometry of the 
worldvolume, $\Gamma^c_{ab} = g^{cd} e_{\mu\,d} \partial_a e^\nu{}_b$, 
when a fixed metric exists in the bulk; this is the case, for 
example, when the Minkowski metric is expressed in spherical 
coordinates. As before, the differentiation of~(\ref{L2}) 
with respect to the gradients of the tangent vectors leads to
\begin{equation}
\label{partial2}
e^\mu{}_r \frac{\partial \mathcal{L}_{(2n)\,\mbox{\tiny bulk}}}{\partial 
(\partial_c e^\mu{}_r)} = -2 \sqrt{-g}\,Q_a{}^{bcd} \Gamma^a_{bd}.
\end{equation}
In arriving to the last equality we have used~(\ref{symm1}) and 
the symmetries of the Christoffel symbols. In this sense, 
comparison with the relation~(\ref{L1}) furnishes us with another 
holographic relationship
\begin{equation}
\mathcal{L}_{(2n)\,\mbox{\tiny surf}} = - \partial_a \left(  
e^\mu{}_b \frac{\partial \mathcal{L}_{(2n)\,\mbox{\tiny bulk}}}{\partial 
(\partial_a e^\mu{}_b)} \right).
\end{equation}
Such transformation leads to express~(\ref{holo1}) as 
\begin{equation}
\mathcal{L}_{(2n)} = \mathcal{L}_{(2n)\,\mbox{\tiny bulk}}
- \partial_a \left( e^\mu{}_b \, \frac{\partial 
\mathcal{L}_{(2n)\,\mbox{\tiny bulk}}}{\partial (\partial_a 
e^\mu{}_b)} \right).
\label{holographic111}
\end{equation}
This particular expression serves to make contact with the 
Hamiltonian framework behind this effective field theory. 
With support with the Ostrogradsky Hamiltonian formalism \cite{nesterenko1989,CapoGuvenRojas2004,capo2017}, the canonical 
momentum conjugate to $e^\mu{}_a = \partial_a X^\mu$ is
\begin{equation}
\label{momentum}
\, _{(2n)}\mathcal{P}^{ab}_\mu := \frac{\partial 
\mathcal{L}_{(2n)\,\mbox{\tiny bulk}}}{\partial (\partial_a 
e^\mu{}_b)},
\end{equation}
so that the relation~(\ref{holographic111}) becomes
\begin{equation}
\mathcal{L}_{(2n)} = \mathcal{L}_{(2n)\,\mbox{\tiny bulk}}
- \partial_a \left( e^\mu{}_b \,\,  _{(2n)}\mathcal{P}^{ab}_\mu 
\right).
\label{holographic1111}
\end{equation}
This is precisely the form of the so-called $``d(q p)$''
structure introduced by Padmanabhan in~\cite{padmanabhan2006,padmanabhan2006b,padmanabhan2002},
which assumes that it is valid for many gravitational theories.
 
Regarding the last alternative, for the RT model when the definition of the 
Christoffel symbols, $\Gamma^a_{bc} = e_{\mu}{}^a \partial_b e^\mu{}_c$,
is introduced into~(\ref{L2}) and after a arrangement of the various terms floating
around, the corresponding bulk Lagrangian density~(\ref{L2}) becomes
\begin{equation}
\mathcal{L}_{\mathcal{R}\,\mbox{\tiny bulk}} = \sqrt{-g}\,
M^{abcd}_{\mu\nu}\,\partial_a e^\mu{}_b \partial_c e^\nu{}_d,
\label{L2b}
\end{equation}
where 
\begin{equation}
M^{abcd}_{\mu\nu} := -2 g^{a[b} e_\mu{}^{|c|} e_\nu{}^{d]}. 
\end{equation}
Notice that~(\ref{L2b}) is now quadratic in $e^\mu{}_a$.

\subsection{Odd case}

In analogy with the analysis performed for the even case, from the 
expression~(\ref{oddL}) it follows that the structure~(\ref{dens1}) holds 
in the odd case. For $n=1,2,3,\ldots$
\begin{eqnarray}
\nonumber
L_{(2n+1)} &= \frac{1}{2^n} \delta^{a_1 a_2 a_3 a_4 \cdots a_{2n} a_{2n+1}}_{b_1 
b_2 b_3 b_4 \cdots b_{2n} b_{2n+1}} \,g^{bb_{2n}} g^{cb_{2n+1}}\,
\mathcal{R}_{a_1 a_2}{}^{b_1b_2} \cdots 
\nonumber
\\
& \qquad \qquad \qquad \qquad \cdots \mathcal{R}_{a_{2n - 3} a_{2n - 2}}{}^{b_{2n-3}b_{2n-2}} 
\mathcal{R}^{b_{2n -1}}{}_{b\,a_{2n-1} a_{2n}} K_{c a_{2n+1}}, 
\nonumber
\\
&= \, _{(2n+1)}\mathsf{Q}_{b_{2n-1}}{}^{a\,a_{2n-1}a_{2n}}\,
\mathcal{R}^{b_{2n -1}}{}_{a\,a_{2n-1} a_{2n}},
\label{eq:oddL}
\end{eqnarray}
where
\begin{eqnarray}
\nonumber
  _{(2n+1)}\mathsf{Q}_{b_{2n-1}}{}^{a\, a_{2n-1} a_{2n}} &:= 
  \frac{1}{2^n} \delta^{a_1 a_2 \cdots a_{2n - 1} 
  a_{2n} a_{2n+1}}_{b_1 b_2 \cdots b_{2n-1} b_{2n} b_{2n+1}} 
  \,g^{a\,b_{2n}} \,\mathcal{R}_{a_1 a_2}{}^{b_1b_2} \cdots
\\
& \qquad \qquad \qquad \cdots \mathcal{R}_{a_{2n-3} a_{2n - 2}}{}^{b_{2n-3} b_{2n-2}} K^{b_{2n+1}}{}_{a_{2n+1}}.
\label{Q2}
\end{eqnarray}
This tensor also inherits the symmetries from the Riemann tensor
\begin{equation}
  _{(2n+1)}\mathsf{Q}^{abcd} = - \,  _{(2n+1)}\mathsf{Q}^{abdc} 
= - \,  _{(2n+1)}\mathsf{Q}^{bacd} = \, _{(2n+1)}\mathsf{Q}^{cdab}. 
\nonumber
\end{equation}
It turns out that~(\ref{Q2}) is constructed from $g^{ab}, 
\mathcal{R}^a{}_{bcd}$ and $K_{ab}$ unlike what happens with 
the even case. A few values of the $\mathsf{Q}$ tensor are
\begin{eqnarray}
\nonumber
 _{(3)}\mathsf{Q}^{abcd} &= J^{a[c}_{(1)} g^{d]b} - g^{a[c} K^{d]b},
\\
\nonumber
 _{(5)}\mathsf{Q}^{abcd} &=  
J^{a[c}_{(3)} g^{d]b} - 3 g^{a[c} J^{d]e}_{(2)} \,K^b{}_e 
+ 6 K^a{}_e \,J^{e[c}_{(1)} K^{d]b} - 3 \mathcal{R}^{aecd}K^b{}_e, 
\end{eqnarray}
where $J^{ab}_{(3)}, J^{ab}_{(2)}$ and $J^{ab}_{(1)}$ are
conserved Lovelock brane tensors. It is therefore possible to write a 
Lagrangian density for the odd case, analogous to~(\ref{dens2}), by writing 
\begin{equation}
\mathcal{L}_{(2n+1)} = \sqrt{-g}\, _{(2n+1)}\mathsf{Q}_{a}{}^{bcd}\,
\mathcal{R}^{a}{}_{bcd}.
\label{dens3}
\end{equation}

In building the splitting of the Lagrangian density for this case into 
a bulk term and a surface term, is clear that the treatment performed for 
the even case, is applied directly. We thus find 
\begin{equation}
\mathcal{L}_{(2n+1)} 
= \mathcal{L}_{(2n+1)\,\mbox{\tiny sur}}  
+ \mathcal{L}_{(2n+1)\,\mbox{\tiny bulk}}
\label{holo22}
\end{equation}
where
\begin{eqnarray}
\mathcal{L}_{(2n+1)\,\mbox{\tiny bulk}} & = 
2 \sqrt{-g}\,  _{(2n+1)}\mathsf{Q}_{a}{}^{b
c d} \Gamma^e _{b c} \Gamma^{a} _{e d},
\\
\mathcal{L}_{(2n+1)\,\mbox{\tiny sur}\,} & = 
\partial_{c}\left( 2\sqrt{-g}\,_{(2n+1)}\mathsf{Q}_{a}{}^{b c d} 
\Gamma^{a}_{b d} \right),
\label{holo3}
\end{eqnarray}
as long as the condition $\nabla_{c} \left[ \, 
_{(2n+1)}\mathsf{Q}_{a}{}^{bcd} \right] =0$
is satisfied.

It must be pointed out that from the functional dependence 
of the variables of the tensor~(\ref{Q2}), $\mathsf{Q}= 
\mathsf{Q} (g^{ab}, \mathcal{R}^a{}_{bcd}, K_{ab})$, it 
immediately follows that the structure of~(\ref{partial}), 
(\ref{holographic1}) and (\ref{holographic11}), holds for the 
odd case. Certainly, the structures given in~(\ref{holographic1}) and~(\ref{holographic11}) are still maintained for this case
\begin{equation}
\label{holographic2}
\mathcal{L}_{(2n+1)\,\mbox{\tiny sur}} = - \partial_a
\left( \delta^c_b \, \frac{\partial 
\mathcal{L}_{(2n+1)\,\mbox{\tiny bulk}}}{\partial \Gamma^c_{ab}}  \right).
\end{equation}
We find then that~(\ref{holo22}), with the aid of the identity~(\ref{holographic2}), 
can be expressed as
\begin{equation}
\mathcal{L}_{(2n+1)} = \mathcal{L}_{(2n+1)\,\mbox{\tiny bulk}}
- \partial_a \left( \delta^c_b \, \frac{\partial 
\mathcal{L}_{(2n+1)\,\mbox{\tiny bulk}}}{\partial \Gamma^c_{ab}} \right),
\label{holographic21}
\end{equation}
thus establishing holographic relationship for the odd case. 

In analogy with the even case, the expression~(\ref{partial2}) remains 
valid in this case as long as  we take into account that the 
tensor~(\ref{Q2}) does not explicitly depend on the gradient 
of the tangent vectors. This serves to establish another holographic 
relationship for the odd case. Additionally, the relationship~(\ref{holographic1111}) is still valid for this 
case so we can also ensure a ``$d(q p)$'' structure inherent in this 
context. The proof that the tensor~(\ref{Q2}) is conserved is 
provided in~\ref{app1b}.

\section{Relationships among conserved Lovelock type brane tensors}

We do not attempt to attribute any sophisticated interpretation, 
either physical or geometrical, to the $Q$ tensors. In turn, we 
believe that tensors $J$ have a clearer geometric interpretation 
since they come from the invariance under reparametrizations of 
the worlvolume. In fact, it turns out that these tensors are 
anchored in some manner.

It follows from the contraction of definition~(\ref{eq:conserved1}) with 
the extrinsic curvature, an expression for the $n$th order Lovelock type
brane invariant 
\begin{equation}
L_{(p+1)} = J^{ab} _{(p)} K_{ab},
\label{eq:relation}
\end{equation}
with $p=0,1,2,\ldots$. With this identity we shall relate tensors $Q$ 
in favor of tensors $J$. Certainly, when the Gauss-Codazzi integrability condition, 
$\mathcal{R}_{abcd} = K_{ac}K_{bd} - K_{ad}K_{bc}$, is inserted 
into the expression~(\ref{eq:evenL}) and by invoking the symmetries 
of the $Q$ tensor, we get
\begin{equation}
L_{(2n)} = 2\, _{(2n)}Q^{abcd} K_{ac} K_{bd}.
\end{equation}
Now, when the expression~(\ref{eq:relation}) enters the game, this 
allows to identify a relationship among the conserved tensors of 
the Lovelock brane gravity theory 
\begin{equation}
J^{ab}_{(2n-1)} = 2\, _{(2n)}Q^{acbd} K_{cd}.
\label{oddJJ}
\end{equation}
It is important to remark that the conservation of the 
tensors~(\ref{Q1}) and~(\ref{Q2}) is due to the invariance 
under reparametrizations of the worldvolume. Knowing that 
$\nabla_b J^{ab}_{(n)} = 0$ must be fulfilled, it is necessary 
that the $Q$ tensors must be conserved, as it happens.

Clearly, a similar procedure with the relations~(\ref{eq:oddL}) 
and~(\ref{eq:relation}), helps to identify another relationship 
among the conserved tensors of the theory 
\begin{equation}
J^{ab}_{(2n)} = 2\, _{(2n+1)}\mathsf{Q}^{acbd}K_{cd}.
\label{evenJJ}
\end{equation}
It is worthwhile to remark that~(\ref{oddJJ}) and~(\ref{evenJJ}) are
valid for $n=1,2,3,\ldots$.

In~\cite{cruzrojas2013,bccrojas2016} was proved that for a $p$th order
Lovelock type brane model, the associated field equation is
\begin{equation}
L_{(p+1)} = J^{ab}_{(p)} K_{ab} = 0.
\label{eom}
\end{equation}
In this sense, from the relations~(\ref{oddJJ}) and~(\ref{evenJJ}), 
it is fairly easy to express the equations of motion in terms 
of the conserved $Q$ tensors. 

For an even Lovelock type brane model, $L_{(2n)}$, the 
equation of motion results $L_{(2n +1)} = J^{ab}_{(2n)} K_{ab}= 0$. 
That is, $2 \, _{(2n+1)}\mathsf{Q}^{acbd} K_{cd} K_{ab}=0$, where 
the expression~(\ref{evenJJ}) has been invoked. Thus, by using
back the Gauss-Codazzi condition, we have that the associated 
equation of motion is
\begin{equation}
\label{eom-even}
\, _{(2n + 1)}\mathsf{Q}_a{}^{bcd}\mathcal{R}^a{}_{bcd} = 0.
\end{equation}
Similarly, for an odd Lovelock type brane model, $L_{(2n + 1)}$, 
the equation of motion is $L_{(2n + 2)} = J^{ab}_{(2n + 1)} K_{ab} 
= 0$. In such a case, by considering~(\ref{oddJJ}), this equation 
of motion also takes a compact form
\begin{equation}
\label{eom-odd}
\, _{(2n + 2)}Q_a{}^{bcd} \mathcal{R}^a{}_{bcd} = 0.
\end{equation}
Doubtless,~(\ref{eom-even}) and~(\ref{eom-odd}) make up a 
rendering of the expression~(\ref{eom}) in terms of the 
Riemann tensor.

Furthermore, in passing we would like to mention the relation
among the conserved stress tensor $f_{(n)}^{a\,\mu}$ for the 
$n$th order LBI with the $Q$ tensors in dependence of the 
nature of $n$,~\cite{cruzrojas2013}. This is quite straightforward 
by considering identities~(\ref{oddJJ}) and~(\ref{evenJJ}). For 
the even case we have
\begin{equation}
\label{faa}
f^{a\,\mu}_{(2n)} = 2\, _{(2n+1)}{Q}^{acbd}K_{cd}\, e^{\mu}{}_b,
\end{equation} 
whereas for the odd case, we get
\begin{equation}
\label{fab}
f^{a\,\mu}_{(2n-1)} = 2\, _{(2n -1)}\mathsf{Q}^{acbd}K_{cd} \, 
e^{\mu}{}_b.
\end{equation} 
What remains to be done is an analysis of both the physical 
utility as well as a deep geometric interpretation of this 
type of holographic relationships.

\section{On the $K$ brane action}

Within the framework behind~(\ref{holo1}) and~(\ref{holo22}), the 
dependence on the curvature tensor has been essential in our development 
when we perform the covariant separation of the Lovelock type brane 
invariants. A glance at~(\ref{oddL}) shows that for $n=1$, the scheme 
adopted here rules out the case of the $K$ brane model because this can 
not be written in terms of the Riemann tensor. This is quite evident 
since this model is linear in the mean extrinsic curvature. Thence, this 
particular case needs a careful handling, or different, since the simple 
expression, $\sqrt{-g} \,K$, does not help to reach the purpose. Indeed, 
the Lagrangian density for this model in presence of a fixed background 
spacetime is
\begin{eqnarray}
\mathcal{L}_K &= \sqrt{-g}\,g^{ab}K_{ab},
\nonumber
\\
&= - \sqrt{-g}\,G_{\mu\nu} n^\mu g^{ab} \partial_a
e^\nu{}_b - \sqrt{-g}\,G_{\mu\nu} n^\mu \Gamma^\nu_{\alpha \beta} 
\mathcal{H}^{\alpha \beta},
\label{LK1}
\end{eqnarray}
where $G_{\mu\nu}$ is the bulk metric, $\Gamma^\mu_{\alpha 
\beta}$ are the background Christoffel symbols and $\mathcal{H}^{\mu\nu}
:= g^{ab} e^\mu{}_a e^\nu{}_b$ is the projection operator in 
$\mathcal{M}$ onto $m$. Although the strategy of identifying a 
surface term seems feasible, here is not directly of much use 
since by considering an integration by parts, and making use 
of the definition of the extrinsic curvature this fails due to 
one faces the orthonormality of the worldvolume basis, $e_a \cdot n =0$.

An alternative way of handling this issue is by breaking the 
worldvolume covariance and thus perform a Hamiltonian analysis 
supported by a geometric ADM framework. Then, inspired by the 
geometric approach made in~\cite{CapoGuvenRojas2000,CapoGuvenRojas2004} 
we can express~(\ref{LK1}) as
\begin{eqnarray}
\mathcal{L}_K &= - \sqrt{-g}\,g^{AB} G_{\mu\nu} n^\mu 
\partial_A \epsilon^\nu{}_B - 2\sqrt{-g}\,\frac{N^A}{N^2} 
\,G_{\mu\nu} n^\mu \,\partial_A \dot{X}^\nu + \frac{\sqrt{h}}{N}
\,G_{\mu\nu}\,n^\mu  \ddot{X}^\nu
\nonumber
\\
& - \sqrt{-g} \,G_{\mu\nu} n^\mu \Gamma^\nu _{\alpha \beta} 
\,\mathcal{H}^{\alpha \beta},
\end{eqnarray}
where $\dot{X}^\mu = \partial_t X^\mu$, $t$ is a coordinate 
that labels the leafs of the foliation of the worldvolume 
by $\Sigma_t$ and $h:= \det (h_{AB})$ with $h_{AB}$ being the 
spacelike metric defined on $\Sigma_t$ such that $\sqrt{-g} =
N \sqrt{h}$. Further, $N$ and $N^A$ are the lapse function and 
the shift vector, respectively. Here is more explicit the linear dependence 
of this Lagrangian density on the accelerations of the brane 
so that this model also is affine in acceleration~\cite{affine2016}. 
A possibility to continue with the computation is to take 
advantage from the fact that for a codimension one worldvolume, 
the normal vector can be written in terms of the velocity of the 
brane, $\dot{X}^\mu$, the Levi-Civita symbols and the geometry of 
the brane $\Sigma_t$ in order to reduce the expression. 
However, a further develop gives rise to a splitting rather 
difficult at this time so that we are unable to immediately 
assess the accuracy of this alternative. Therefore, this issue 
needs to be further analyzed before to provide a total conclusion. 
We will report this development elsewhere.

\section{Concluding remarks}
\label{sec:6}

We have shown that Lovelock type brane gravity 
has holographic relationships similar in spirit to those of 
pure Lanczos-Lovelock gravity. These relationships allow us to 
extract information about a part of the splitted Lagrangian 
density, in terms of the other one. A distinctive feature of this 
framework is the strong dependence on the conservation condition 
on the $Q$ tensors, (\ref{Q1}) and (\ref{Q2}). As shown, this is 
a consequence of the invariance under reparametrizations of the 
action functional in this type of gravity.  In each surface 
governed by this action, there is only one degree of freedom, 
corresponding to the breathing mode of the worldvolume so its 
nature is only geometric~\cite{bccrojas2016}. In this sense, it 
is expected that the Lagrangians, either $\mathcal{L}_{\mbox{\tiny bulk}}$ 
or $\mathcal{L}_{\mbox{\tiny surf}}$ reflect this fact, thereby 
encoding the same amount of dynamical content by describing the 
same degree of freedom. The next step in this development is to 
explore the physical consequences coming from the holographic 
relationships. We expect to show eventually how these relationships 
get reflected at the boundary surface of the worldvolumes governed 
by this gravity. Furthemore, the holographic relationships that 
were found here also preserve the so-called ``$d (qp)$''
structure~\cite{padmanabhan2006,padmanabhan2006b}, which can be 
thought of as a part of the transition from a coordinate representation 
to a momentum representation in dependence of the chosen variables.  
Additionally, in order to continue the quantum analysis for 
these brane models, the Hamiltonian formalism can benefit from 
this approach by considering only the Lagrangian $\mathcal{L}_{\mbox{\tiny 
bulk}}$ since the second-order nature of the full Lagrangian density 
is encoded in the Lagrangian $\mathcal{L}_{\mbox{\tiny surf}}$ thus avoiding 
a cumbersome Ostrogradski Hamiltonian approach.

\section*{Acknowledgments}
I am grateful to R. Capovilla, R. Cordero, M. Cruz and C. Campuzano for 
the stimulating discussions and valuable comments. The author also thanks 
ProdeP-M\'exico, CA-UV-320: Algebra, Geometr\'\i a y Gravitaci\'on. 
This work was partially supported by SNI (M\'exico).

\appendix

\section{Proofs of the divergenless of the $Q$ terms}
\label{app1}

In this Appendix we provide the proofs of the divergenless property of 
the tensors~(\ref{Q1} and~(\ref{Q2}).

\subsection{Proof of $\nabla_{c} \left[ _{(2n)}Q_{a}{}^{bcd} \right] =0$.}
\label{app1a}

From Eq.~(\ref{Q1}) we have
\begin{eqnarray}
\fl \nabla_c \left[ _{(2n)}Q_a{}^{bcd} \right] &=  \frac{1}{2^n} \delta^{a_1 a_2 \cdots 
a_{2n -3} a_{2n-2} \,c d}_{b_1 b_2 \cdots b_{2n-3} b_{2n - 2} \,a b_{2n}}
g^{b b_{2n}} \left[ \nabla_c \mathcal{R}_{a_1 a_2}{}^{b_1 b_2} \mathcal{R}_{a_3 a_4}{}^{b_3 b_4} 
\cdots \mathcal{R}_{a_{2n-3} a_{2n-2}}{}^{b_{2n-3} b_{2n - 2}} \right. 
\nonumber
\\
\fl & \left. \quad \qquad \qquad \qquad + \cdots + \mathcal{R}_{a_1 a_2}{}^{b_1 b_2} 
\mathcal{R}_{a_3 a_4}{}^{b_3 b_4} 
\cdots \nabla_c \mathcal{R}_{a_{2n-3} a_{2n-2}}{}^{b_{2n-3} b_{2n - 2}} \right],
\nonumber
\\
&=  \frac{(n-1)}{2^n} \delta^{a_1 a_2 \cdots a_{2n -3} a_{2n-2} \,c d}_{b_1 b_2 \cdots 
b_{2n-3} b_{2n - 2} \,a b_{2n}} g^{b b_{2n}} \, \mathcal{R}_{a_1 a_2}{}^{b_1 b_2} 
\mathcal{R}_{a_3 a_4}{}^{b_3 b_4} \cdots 
\nonumber
\\
\fl & \qquad \qquad \qquad \qquad \qquad\qquad \qquad \qquad \quad \cdots 
\nabla_c \mathcal{R}_{a_{2n-3} a_{2n-2}}{}^{b_{2n-3} b_{2n - 2}}.
\nonumber
\end{eqnarray}
In view of the antisymmetry of the gKd it follows the relation
\begin{eqnarray}
 \nabla_c \left[ _{(2n)}Q_a{}^{bcd} \right] &= \frac{(n-1)}{2^n} \delta^{a_1 a_2 \cdots 
a_{2n -3} a_{2n-2} \,c d}_{b_1 b_2 \cdots 
b_{2n-3} b_{2n - 2} \,a b_{2n}} g^{b b_{2n}} \, \mathcal{R}_{a_1 a_2}{}^{b_1 b_2} 
\mathcal{R}_{a_3 a_4}{}^{b_3 b_4} \cdots 
\nonumber
\\
 &  \qquad \qquad \qquad\qquad \qquad \qquad \quad \cdots 
\nabla_{[c} \mathcal{R}_{a_{2n-3} a_{2n-2}]}{}^{b_{2n-3} b_{2n - 2}}.
\nonumber
\end{eqnarray}
From the usual Bianchy identity, $\nabla_{[a} \mathcal{R}_{bc]de} = 0$, we infer
that
\begin{equation}
 \nabla_{c} \left[ _{(2n)}Q_{a}{}^{bcd} \right] =0,
\end{equation}
so that, the assertion is proved.

\subsection{Proof of $\nabla_{c} \left[\, _{(2n+1)}\mathsf{Q}_{a}{}^{bcd} \right] =0$.}
\label{app1b}

From Eq.~(\ref{Q2}) we have
\begin{eqnarray}
\nabla_c \left[ \, _{(2n+1)}\mathsf{Q}_a{}^{bcd} \right] &=  \frac{1}{2^n} \delta^{a_1 a_2 \cdots 
a_{2n-3} a_{2n-2} \, c\,d\, a_{2n+1}}_{b_1 b_2 \cdots b_{2n-3} b_{2n -2}\, a\, b_{2n} b_{2n+1}}
g^{b b_{2n}} g^{e b_{2n+1}} \left( \nabla_c \mathcal{R}_{a_1 a_2}{}^{b_1 b_2} 
\cdots \right.
\nonumber
\\
 & \left. \qquad \qquad \qquad \qquad \qquad
\cdots \mathcal{R}_{a_{2n-3} a_{2n-2}}{}^{b_{2n-3} b_{2n - 2}} K_{e\,a_{2n+1}}  \right. 
\nonumber
\\
 & \left. + \cdots \mathcal{R}_{a_1 a_2}{}^{b_1 b_2} \cdots 
\nabla_c \mathcal{R}_{a_{2n-3} a_{2n-2}}{}^{b_{2n-3} b_{2n - 2}} K_{e\,a_{2n+1}} \right. 
\nonumber
\\
& \left. + \mathcal{R}_{a_1 a_2}{}^{b_1 b_2} \cdots 
\mathcal{R}_{a_{2n-3} a_{2n-2}}{}^{b_{2n-3} b_{2n - 2}} \nabla_c K_{e\,a_{2n+1}} \right)
\nonumber
\\
& = \frac{(n-1)}{2^n} \delta^{a_1 a_2 \cdots 
a_{2n-3} a_{2n-2} \, c\,d\, a_{2n+1}}_{b_1 b_2 \cdots b_{2n-3} b_{2n -2}\, a\, b_{2n} b_{2n+1}}
g^{b b_{2n}} g^{e b_{2n+1}} \mathcal{R}_{a_1 a_2}{}^{b_1 b_2} \cdots
\nonumber
\\
& \qquad \qquad \qquad \qquad \quad \cdots \nabla_c \mathcal{R}_{a_{2n-3} a_{2n-2}}{}^{b_{2n-3} b_{2n - 2}} 
K_{e\,a_{2n+1}} 
\nonumber
\\
& + \frac{1}{2^n} \delta^{a_1 a_2 \cdots 
a_{2n-3} a_{2n-2} \, c\,d\, a_{2n+1}}_{b_1 b_2 \cdots b_{2n-3} b_{2n -2}\, a\, b_{2n} b_{2n+1}}
g^{b b_{2n}} g^{e b_{2n+1}} \mathcal{R}_{a_1 a_2}{}^{b_1 b_2} \cdots
\nonumber
\\
& \qquad \qquad \qquad \qquad \quad \cdots \mathcal{R}_{a_{2n-3} a_{2n-2}}{}^{b_{2n-3} b_{2n - 2}} 
\nabla_c K_{e\,a_{2n+1}} 
\nonumber
\end{eqnarray}
As before, in view of the antisymmetry of the gKd as well the Codazzi-Mainardi 
integrability condition for extended objects of arbitrary dimensions in Minkowski 
spacetime, $\nabla_a K_{bc} = \nabla_b K_{ac}$, we have
\begin{eqnarray}
\nabla_c \left[ \, _{(2n+1)}\mathsf{Q}_a{}^{bcd} \right] &=
\frac{(n-1)}{2^n} \delta^{a_1 a_2 \cdots 
a_{2n-3} a_{2n-2} \, c\,d\, a_{2n+1}}_{b_1 b_2 \cdots b_{2n-3} b_{2n -2}\, a\, b_{2n} b_{2n+1}}
g^{b b_{2n}} g^{e b_{2n+1}} \mathcal{R}_{a_1 a_2}{}^{b_1 b_2} \cdots
\nonumber
\\
& \qquad \qquad \qquad \qquad \quad \cdots \nabla_{[c} \mathcal{R}_{a_{2n-3} a_{2n-2}]}{}^{b_{2n-3} b_{2n - 2}} 
K_{e\,a_{2n+1}} 
\nonumber
\\
& + \frac{1}{2^n} \delta^{a_1 a_2 \cdots 
a_{2n-3} a_{2n-2} \, c\,d\, a_{2n+1}}_{b_1 b_2 \cdots b_{2n-3} b_{2n -2}\, a\, b_{2n} b_{2n+1}}
g^{b b_{2n}} g^{e b_{2n+1}} \mathcal{R}_{a_1 a_2}{}^{b_1 b_2} \cdots
\nonumber
\\
& \qquad \qquad \qquad \qquad \quad \cdots \mathcal{R}_{a_{2n-3} a_{2n-2}}{}^{b_{2n-3} b_{2n - 2}} 
\nabla_e K_{c\,a_{2n+1}} 
\nonumber
\end{eqnarray}
By invoking the Bianchy identity, $\nabla_{[a} \mathcal{R}_{bc]de} = 0$, as 
well the antisymmetry of the gKd and the symmetry of the extrinsic curvature, 
it follows that
\begin{equation}
\nabla_{c} \left[ \,_{(2n+1)}\mathsf{Q}_{a}{}^{bcd} \right] =0.
\end{equation}
This proves our assertion.

\bigskip
\noindent
\textbf{References}

\bigskip


\end{document}